\documentclass[jkps,twocolumn,showpacs,showkeys]{revtex4}
\usepackage[pdftex]{graphicx}
\usepackage{amssymb}
\usepackage{amsmath}
\usepackage{bm}
\usepackage{xcolor}
\usepackage[normalem]{ulem}

\begin{document}
\title[]{Thickness dependent charge density wave networks on thin $1T$-TaS$_2$}

\author{Wooin \surname{Yang}}
\author{Dowook \surname{Kim}}
\author{Hyoung Kug \surname{Kim}}
\author{Tae-Hwan \surname{Kim}}
\email{taehwan@postech.ac.kr}
\affiliation{Department of Physics, Pohang University of Science and Technology (POSTECH), Pohang 37673, Korea}

\date[]{Received }

\begin{abstract}
We investigate mechanically exfoliated thin $1T$-TaS$_2$ with scanning tunneling microscopy at room temperature.
Sample preparation without air exposure enables access to intrinsic charge-density-wave (CDW) phases of thin $1T$-TaS$_2$.
At room temperature, we can observe the expected nearly commensurate CDW (NCCDW) phase on thin flakes similar to bulk $1T$-TaS$_2$.
Further analysis reveals that the CDW domains in the NCCDW phase become smaller and have more anisotropic shape with decreasing thickness in the range of 8--28 layers.
Our findings demonstrate that the anisotropic CDW nature of thin $1T$-TaS$_2$ would be crucial to understand its exotic CDW-related phenomena
and demand a systematic study on its correlation between the thickness-driven CDW domain anisotropy and the intermediate CDW states in thin $1T$-TaS$_2$. 
\end{abstract}

\pacs{68.37.Ef, 68.55.Jk, 68.35.Rh, 71.30.+h}

\keywords{scanning tunneling microscopy, $1T$-TaS$_2$, charge density wave, exfoliation, thickness dependence}

\pagenumbering{roman} 

\maketitle

\clearpage
\pagenumbering{arabic} 

\section{Introduction}

Two-dimensional (2D) layered materials enable intensive research on novel quantum phases of condensed matters due to their unique extraordinary electrical, optical, and magnetic properties~\cite{Novoselov2016}.
Such exotic 2D materials can exhibit not only rich 2D physics but also potential possibilities for future applications in quantum devices.
Furthermore, mechanical exfoliation has been widely used to produce atomically thin 2D materials since the discovery of graphene. 
Convenient thickness control of 2D layered materials via the exfoliation opens a new way to realize high-performance electronic and optical devices without physical and chemical damage. 

Among the emerging 2D materials, $1T$-TaS$_2$, one of widely studied layered transition metal dichalcogenides, provides an unprecedented opportunity to study intertwined quantum phases such as a Mott insulating state, peculiar charge-density-wave (CDW) phases, quantum spin liquid states, and superconductivity~\cite{Law2017, Qiao2017, Gerasimenko2019, Zhu2019, Butler2020, Sipos2008}.
In addition, the intricate interplay between quantum phases in $1T$-TaS$_2$ has been explored by thinning thickness via several mechanical exfoliation techniques~\cite{Yoshida2014, Yu2015, Hollander2015, Tsen2015, Vaskivskyi2016}.
In bulk, metallic $1T$-TaS$_2$ exhibits an incommensurate CDW (ICCDW) phase below 550~K and a nearly commensurate CDW (NCCDW) below 350~K.
The NCCDW phase consists of a quasi-hexagonal array of commensurate domains with incommensurate metallic domain wall networks.
With decreasing temperature, commensurate domains gradually expand and eventually the CDW phase is fully commensurate with respect to the underlying lattice below 180~K~\cite{Scruby1975, Wu1990, Nakatsugawa2020}.
In the commensurate CDW (CCDW) phase or the commensurate domains in NCCDW, each unit cell forms a $(\sqrt{13}\times\sqrt{13})R13.9\,^{\circ}$ reconstruction with 13 Ta atoms, leading to the Star of David (SD) building block. 

Such CDW phases of $1T$-TaS$_2$ have been reported to be extremely sensitive to external stimuli including temperature, pressure, electrical bias, optical illumination, and chemical doping~\cite{Sipos2008, Stojchevska2014, Vaskivskyi2015, Ang2015, Vaskivskyi2016}. 
For example, some $1T$-TaS$_2$ devices demonstrated that the CDW phase transition from NCCDW to ICCDW has the strong potential for oscillators, detectors, and radiation-hard electronics even at room temperature~\cite{Liu2016, Liu2017, Khitun2017, Khitun2018, Taheri2022}.
Though the phase transitions have often been interpreted as inhomogeneous mixtures between stable quantum phases, detailed understanding of the interplay between quantum phases are still lacking~\cite{Yoshida2015, Yoshida2017, Patel2020}.
Furthermore, many experimental works have focused on the phase transition between CCDW and NCCDW phases in bulk $1T$-TaS$_2$.
However, there are a few limited studies investigating the NCCDW–ICCDW phase transition on thin $1T$-TaS$_2$ at room temperature~\cite{Ishiguro2020}, which can provide more compelling opportunities for realizing device applications using the CDW phase transition~\cite{Taheri2022}. 

Here, we have studied pristine CDW phase of thin $1T$-TaS$_2$ in real space by scanning tunneling microscopy (STM) to investigate the correlation between thickness and the CDW phase at room temperature. 
We observed the expected NCCDW phase on thin flakes similar to bulk $1T$-TaS$_2$.
Moreover, we found that CDW domains become smaller and have a more anisotropic structure with decreasing thickness.
Our finding indicates that the anisotropic CDW nature of thin $1T$-TaS$_2$ would be important to understand its exotic CDW-related phenomena
and to develop device applications at room temperature. 
\begin{figure}[ht]
\begin{center}
\includegraphics[width=\linewidth]{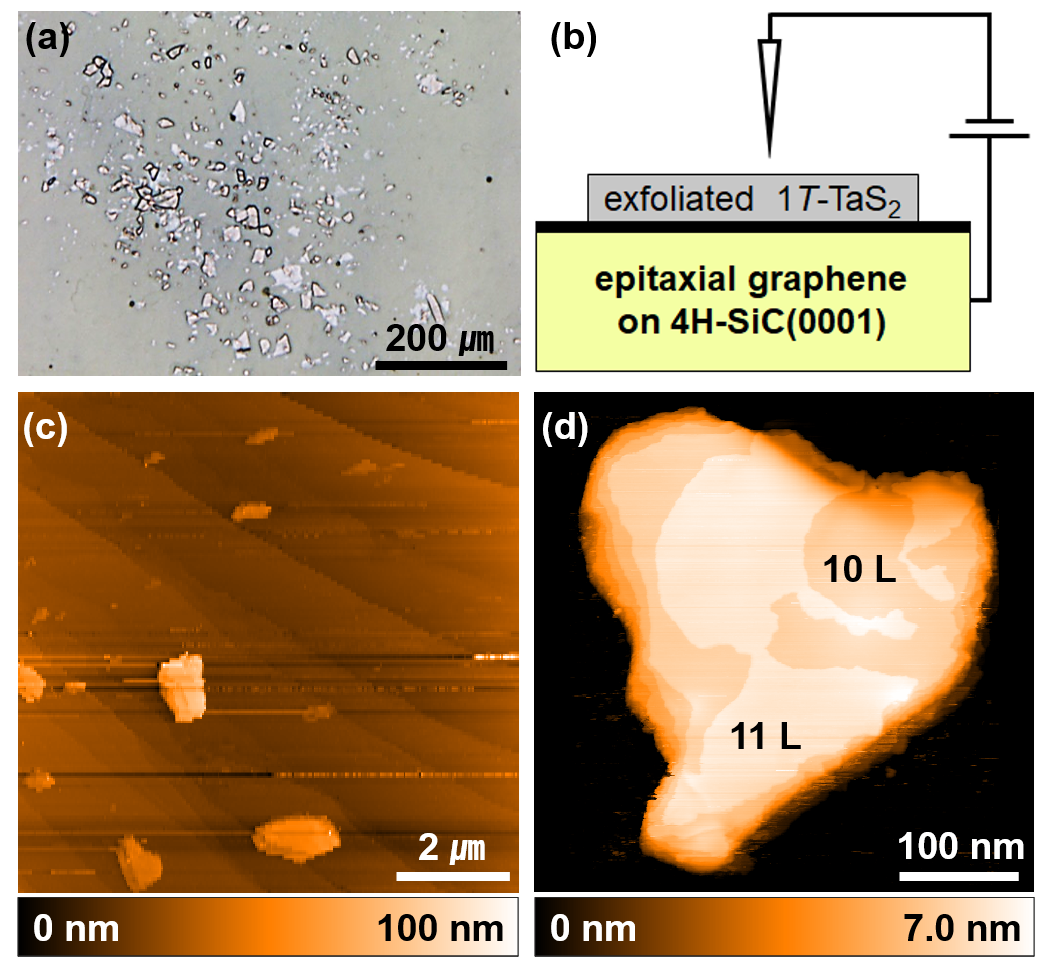}
\caption{Scanning tunnelinng microscopy (STM) measurements on mechanically exfoliated thin $1T$-TaS$_2$ flakes.
(a) Optical microscopy image after transferring exfoliated thin $1T$-TaS$_2$ flakes onto epitaxial graphene grown on a $4H$-SiC(0001) single crystalline substrate. 
(b) Schematic drawing of our STM experimental setup. 
(c) Large-scale STM topographic image showing several thin $1T$-TaS$_2$ flakes with varying thicknesses. 
(d) Typical STM image of a $1T$-TaS$_2$ flake showing mainly 10~L and 11~L.
}
\end{center}
\end{figure}

\section{Experimental}

To obtain thin $1T$-TaS$_2$ flakes with different thicknesses, we used mechanical exfoliation of bulk single crystals (purchased from 2D Semiconductors Inc.)~\cite{Kim2016,Park2021a}.
The flakes were repeatedly cleaved with a new residue-free tape in order to make them thin enough down to several layers.
Subsequently, thin flakes were transferred onto graphene, epitaxially grown on a $4H$-SiC(0001) substrate.
The entire sample preparation process has been carried out in an Ar-filled glove box (H$_2$O~$<$~0.1~ppm, O$_2$~$<$~0.1~ppm) to avoid surface oxidation or contamination due to air exposure~\cite{Yamaguchi1997, Shen2020}.
Once the samples have been prepared, we searched for thin enough flakes  by measuring optical transmission contrast with an optical microscope (OM) [Fig.~1(a)].
These OM images were later employed for locating thin flakes in our STM setup.
Then, the samples were carried by a home-built suitcase to an ultrahigh vacuum STM chamber ($P< 1\times 10^{-8}$~Pa) for STM measurements without any exposure to air to preserve clean surfaces of exfoliated $1T$-TaS$_2$ flakes.

Figure~1(b) illustrates our STM experimental setup.
In the STM chamber, we located an electrochemically etched W tip on a target $1T$-TaS$_2$ flake by using an optical navigation system equipped with a long-working-distance OM.
Once we located properly by using OM images, we easily found thin flakes with varying thicknesses within the STM scan range [Fig.~1(c)].
Figure~1(d) shows a representative $1T$-TaS$_2$ flake having 10--11 layers (L). 
We could obtain thin $1T$-TaS$_2$ flakes down to 8~L via our exfoliation method. 
STM images were obtained in a constant-current mode at room temperature where bulk $1T$-TaS$_2$ shows the NCCDW phase.
We typically used a bias voltage ($V_{\rm b}$) of $\pm$0.1--1~V and a tunneling current ($I_{\rm t}$) of 10--100~pA.
To quantitatively analyze observed CDW phases on thin $1T$-TaS$_2$, we applied an affine transform algorithm with respect to atomic Bragg peaks~\cite{Gerasimenko2019a}.
This simple but useful correction can compensate unwanted spatial distortion of STM images due to thermal drift during STM measurements.

\begin{figure}[ht]
\begin{center}
\includegraphics[width=\linewidth]{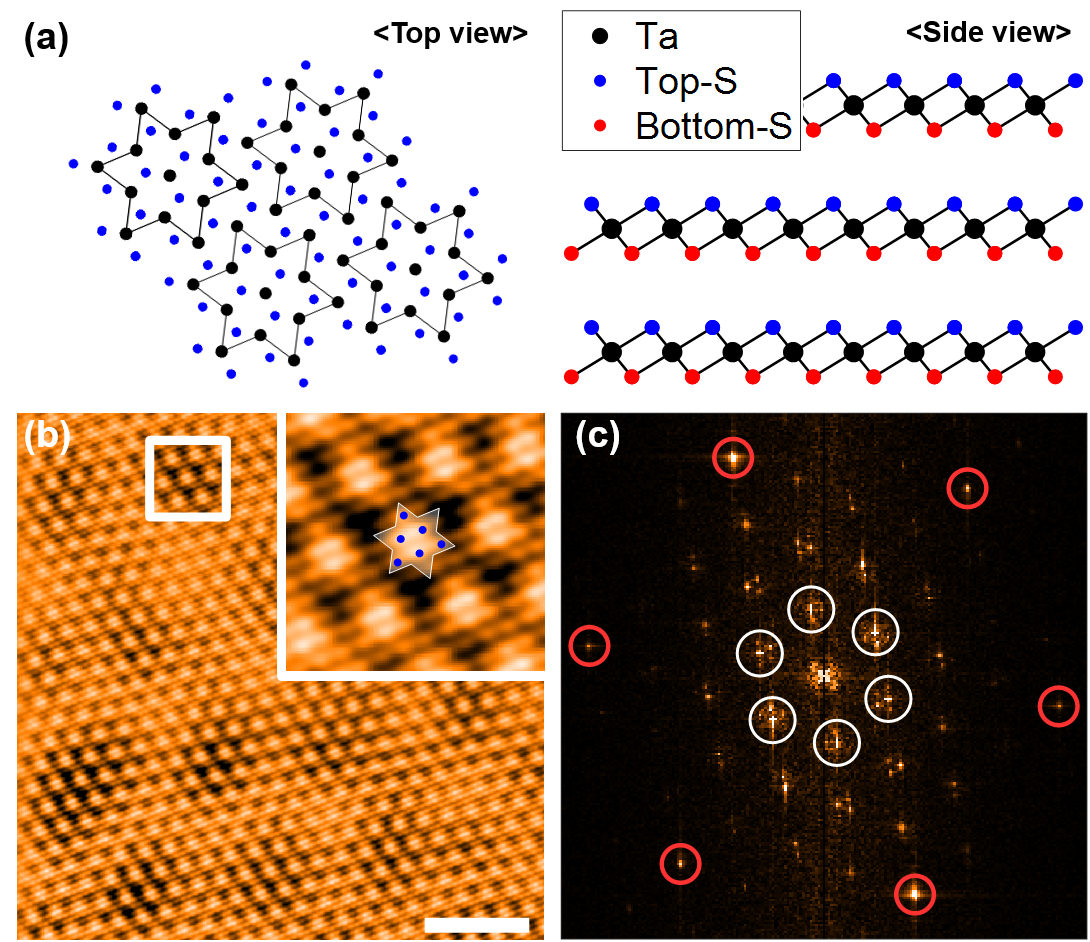}
\caption{Nearly commensurate charge-density-wave (NCCDW) phase of thin $1T$-TaS$_2$. 
(a) Atomic schematic of commensurate CDW (CCDW) in $1T$-TaS$_2$. 
Black, blue, and red dots represent Ta, top S, bottom S atoms, respectively.
For clarity, the bottom S atoms are omitted on the top view.
Each David star with 13 Ta atoms shows the $\sqrt{13} \times \sqrt{13}$ structure of CCDW with respect to the underlying lattice.
For clarity, bottom S atoms are omitted from top view. 
(b) Atom-resolved STM image on 8~L thick $1T$-TaS$_2$ obtained at room temperature ($V_{\rm b}=1$~V, $I_{\rm t}=100$~pA, scale bar = 5~nm). 
A NCCDW phase shows quasi-hexagonal domain networks consisting of CCDW domains and incommensurate domain walls. 
Inset highlights one of CCDW domains indicated by a square. 
(c) Fast Fourier transform of (b). 
Red and white circles denote atomic Bragg points and CDW peaks with their satellites, respectively.
}
\end{center}
\end{figure}



%
%
%
\section{Results and discussion}
Figure~2(a) shows pristine $1T$-TaS$_2$ with a triangular atomic lattice of Ta atoms (indicated by black dots) sandwiched by S atoms (by blue and red dots) in an octahedral coordination. 
In bulk $1T$-TaS$_2$, every unit cell having 13 Ta atoms form a SD CDW cluster (highlighted by black lines) in the CCDW phase below 180~K.
The SD CDW cluster shows the $\sqrt{13} \times \sqrt{13}$ superstructure with respect to the underlying pristine lattice.
On the other hand, the SD CDW cluster appears inside the commensurate domains of the NCCDW phase at higher temperatures (180~K $<T<$ 350~K) in bulk $1T$-TaS$_2$. 
In the NCCDW phase, quasi-hexagonal CCDW domains of the commensurate SD CDW clusters are separated by incommensurate metallic domain walls at room temperature. 

We have investigated exfoliated $1T$-TaS$_2$ flakes with STM at room temperature. 
Figure~2(b) shows an atom-resolved STM topographic image obtained on an 8~L-thick flake.
The STM image clearly reveals quasi-hexagonal commensurate CDW domains separated by incommensurate domain wall networks similar to bulk $1T$-TaS$_2$. 
The zoom-in STM image demonstrates that SD clusters within each CCDW domain exhibit the $\sqrt{13} \times \sqrt{13}$ superstructure of topmost S atoms with respect to the underlying lattice as shown in the inset of Fig.~2(b).
Before analyzing further, we obtained a 2D fast Fourier transform (FFT) from a real space STM image and applied the affine transformation algorithm to correct lattice distortion due to thermal drift.

As shown in Fig.~2(c), the characteristic quasi-hexagonal domain structure is also confirmed by CDW peaks accompanying hexagonal satellite domain peaks (indicated by white circles) with respect to the atomic Bragg peaks (by red circles).
In order to quantify CDW features, an overall CDW periodicity is estimated from the drift-corrected FFT image by averaging the measured distances of three different CDW peaks in the reciprocal space.
An average distance in the reciprocal space corresponds to a characteristic overall CDW period in real space.
For the 8~L flake, we obtain a CDW period of 1.17~nm from Fig.~2(c), which is similar to the previously reported value of bulk $1T$-TaS$_2$ at room temperature~\cite{Singh2022}. 
Based on our observation on the 8~L flake, we can conclude that thin $1T$-TaS$_2$ exhibits more or less the same CDW feature at room temperature as bulk $1T$-TaS$_2$ even though the strong thickness dependence has been reported in the various other experimental measurements~\cite{Yu2015,Yoshida2015,Fu2016}.

\begin{figure}[ht]
\begin{center}
\includegraphics[width=\linewidth]{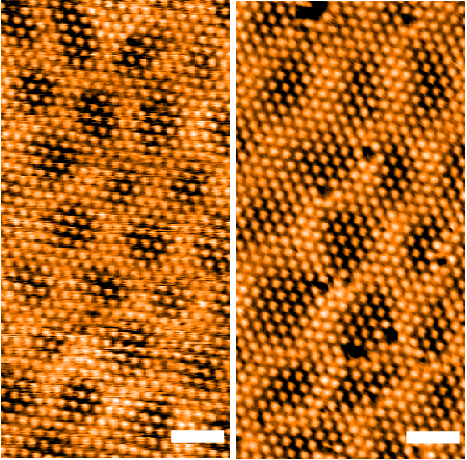}
\caption{Comparison of NCCDW structures between 8~L (left) and 11~L (right). 
A thinner flake shows smaller CDW domains with more anisotropic shape. 
Imaging condition (left/right): $V_{\rm b}$ = $-0.1$/$+0.5$~V, $I_{\rm t}$ = 100/50~pA. Scale bar = 5~nm. 
}
\end{center}
\end{figure}

To see any thickness dependent CDW phases on thin $1T$-TaS$_2$ flakes, we have explored flakes with different thicknesses (8--28~L).
Apparently, we notice that all flakes show the same NCCDW phases at room temperature just like bulk $1T$-TaS$_2$.
However, our careful analysis reveals that thinner flakes have a smaller CDW domain size in the NCCDW phase than thicker ones.
Figure~3 show STM topographic images of 8~L (left) and 11~L (right) $1T$-TaS$_2$ flakes, respectively.
These STM images clearly demonstrate that the thinner 8~L flake significantly has a smaller size of CDW domains than one on the 11~L.
In addition, we found that the thicker flake shows more isotropic domains while thinner one exhibits elongated domains.
To quantify the overall size and anisotropic shape of CDW domains, we analyze the drift-corrected FFT images.
The overall domain size is estimated by the averaged reciprocal distance between CDW peaks and their satellites in FFT images. 
On the other hand, the domain anisotropy is extracted from the ratio of the longest to the shortest domain sizes along three domain orientations.

\begin{figure}[ht]
\begin{center}
\includegraphics[width=\linewidth]{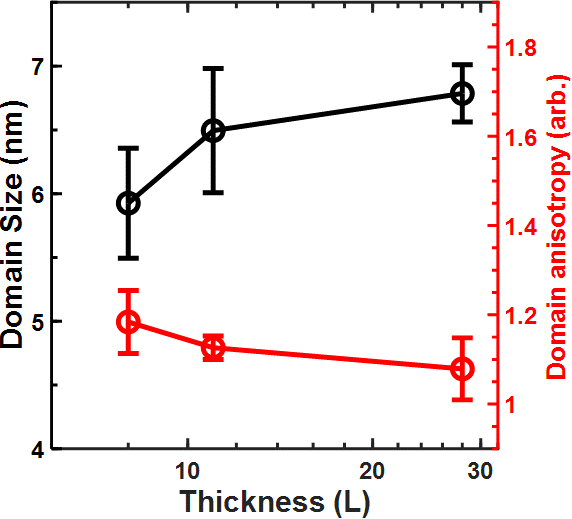}
\caption{Thickness dependent CDW domain size and domain anisotropy of thin $1T$-TaS$_2$.
The overall domain size is estimated by the averaged reciprocal distance between CDW peaks and their satellites in FFT images.
The domain anisotropy is extracted from the ratio of maximum to minimum reciprocal distances between a CDW peak and its satellite peaks in FFT images.
}
\end{center}
\end{figure}

To improve the reliability, we statistically average the domain size and anisotropy of CDW domains as a function of thickness from a sufficient number of atom-resolved STM images on different locations for each thickness.
Figure~4 indicates the domain size and anisotropy of CDW domains as a function of thickness from 8~L to 28~L.
With increasing thickness, the domains become bigger and more isotropic.
Note that different flakes with the same thickness are found to exhibit more or less similar domain sizes.

In bulk $1T$-TaS$_2$, with increasing temperature, the surface areal fraction of the domain walls increases as domains gradually shrink in order to minimize the total energy for forming domains at a given temperature~\cite{Wu1990,Thomson1994,Ishiguro1991}. 
Such tendency of the CDW domain size in bulk with increasing temperature is quite similar to that in thin $1T$-TaS$_2$ flakes with decreasing thickness. 
%
%
This observation strongly suggests that thinning thickness may disturb CDW phases just like thermal heating in bulk.
In this sense, we may effectively consider thinner flakes as bulk $1T$-TaS$_2$ at higher temperatures. 
In other words, such a subtle but distinct trend may be understood by thickness-dependent thermal stability.
This conjecture is also consistent with the other experimental observation that thin $1T$-TaS$_2$ flakes show the lower transition temperature from NCCDW to CCDW with decreasing thickness~\cite{Tsen2015,Yoshida2015,Yu2015,He2016}. 
Thus, thinner flakes with lower thermal stability naturally exhibit smaller and anisotropic commensurate CDW domains. 

%
%
%
%

%
%
%
%
%

\section{Summary}

We systematically investigated mechanically exfoliated thin $1T$-TaS$_2$ in real space with STM at room temperature.
We observed the expected NCCDW phase on all thin flakes similar to bulk $1T$-TaS$_2$.
However, we found the unexpected correlation between thickness and the NCCDW phase.
Further quantitative analysis revealed that the CDW domains become smaller and more anisotropic with decreasing thickness in the range of 8--28~L.
Our finding suggests that thin $1T$-TaS$_2$ would have the thickness-dependent thermal stability in contrast to the bulk counterpart,
which may be important to understand its exotic CDW-related phenomena in thin $1T$-TaS$_2$.
Our observation also demands a systematic study on its correlation between the thickness-driven CDW domain anisotropy and the intermediate CDW states in thin $1T$-TaS$_2$. 



\begin{acknowledgments}
This work was supported by the National Research Foundation of Korea (NRF) funded by the Ministry of Science and ICT, South Korea (Grants No. NRF-2017R1A2B4007742, 2021R1A6A1A10042944, 2021R1F1A1063263, and 2022M3H4A1A04074153).
\end{acknowledgments}


\end{document}